\documentclass[aps,prl,twocolumn,preprintnumbers,superscriptaddress,dblfloatfix,nofootinbib]{revtex4-1}
\usepackage[utf8]{inputenc}
\usepackage{lmodern}
\usepackage{amsmath,amssymb,mhchem}
\usepackage{comment}
\usepackage{graphicx}
\usepackage{url}
\usepackage{color}
\usepackage{enumitem}
\usepackage{subfigure}
\usepackage[dvipsnames]{xcolor}
\usepackage[colorlinks=true,breaklinks=true]{hyperref}
\hypersetup{allcolors=[rgb]{0.0 0.0 0.6},linkcolor=[rgb]{0.75 0.05 0.05}}
\usepackage{bm}
\usepackage{epsfig}
\usepackage{amsmath}
\usepackage{amssymb}
\usepackage{slashed}
\usepackage{color}
\usepackage{accents}
\usepackage{url}
\usepackage[dvipsnames]{xcolor}

\newcommand{\exclude}[1]{}

\begin{document}

\title{Gravitational waves from axion-like particle cosmic string-wall networks}

\author{Graciela B. Gelmini}\email{gelmini@physics.ucla.edu}
\affiliation{Department of Physics and Astronomy, University of California, Los Angeles \\ Los Angeles, California, 90095-1547, USA}

\author{Anna Simpson}\email{ansimps@g.ucla.edu}
\affiliation{Department of Physics and Astronomy, University of California, Los Angeles \\ Los Angeles, California, 90095-1547, USA}

\author{Edoardo Vitagliano}\email{edoardo@physics.ucla.edu}
\affiliation{Department of Physics and Astronomy, University of California, Los Angeles \\ Los Angeles, California, 90095-1547, USA}

\date{\today}

\begin{abstract}
Axion-like particles (ALPs) are a compelling candidate for dark matter (DM), whose production is associated with the formation of a string-wall network. If walls bounded by strings persist, which requires the potential to have multiple local minima ($N>1$), they must annihilate before they become dominant. They annihilate mostly into gravitational waves and non-relativistic ALPs. We show that for ALPs other than the QCD axion these gravitational waves, if produced at temperatures below 
100 eV, could be detected by future cosmological probes for ALPs with mass from $10^{-16}$ to $10^{6}$~eV that could constitute the entirety of the DM.

\end{abstract}
\maketitle
\setcounter{equation}{0}
\setcounter{figure}{0}
\setcounter{table}{0}
\setcounter{section}{0}
\setcounter{page}{1}
 
\section{Introduction}
Gravitational waves (GWs) constitute a powerful tool to assess particle physics models~\cite{Vilenkin:2000jqa,Maggiore:1900zz,Maggiore:2018sht,Sathyaprakash:2009xs,Barack:2018yly}. They could test  in the near future a particular class of light bosonic dark matter (DM) candidates. The production of these particles implies the existence of a stochastic GW background with a peaked spectrum that can be probed by cosmic microwave background (CMB) experiments and astrometry measurements in the  $10^{-16}- 10^{-14}~\mathrm{Hz}$ GW frequency range.

Many extensions of the Standard Model (SM) of elementary particles contain a global $U(1)$ symmetry spontaneously broken at an energy scale $V$ and explicitly broken at another scale $v \ll V$. Models for
the original axion~\cite{Peccei:1977hh, Weinberg:1977ma, Wilczek:1977pj}, invisible axions (also called ``QCD~axions")~\cite{Kim:1979if,Shifman:1979if,Dine:1981rt,Zhitnitsky:1980tq}, and axion-like particles (ALPs) (e.g.~\cite{Svrcek:2006yi,Arvanitaki:2009fg,Acharya:2010zx,Dine:2010cr,Jaeckel:2010ni}) are of this type. 
In these models, the Nambu-Goldstone (NG) boson corresponding to the spontaneous $U(1)$ breaking acquires a mass $m_a \simeq v^2/V$, becoming a pseudo-NG boson
which we denote with $a$ and call an ALP.

If the spontaneous symmetry breaking happens after inflation, as we assume here,  a system of cosmic walls bounded by strings is produced (see e.g. Ref.~\cite{Vilenkin:1984ib} and references therein). Global cosmic strings are created during the spontaneous symmetry breaking, and become connected by walls at a later time $t \simeq m_a^{-1}$.
After the explicit breaking, the potential may have just one minimum, $N=1$, or several, $N > 1$. With $N=1$,
``ribbons" of walls bounded by strings 
surrounded by true vacuum form, which
shrink due to the pull of the walls on the strings. Thus, the wall-string system decays immediately after wall formation, leading to GWs produced only by strings before walls form observable in future pulsar arrays and direct detection experiments if $V\gtrsim 10^{14}~\rm{GeV}$~\cite{Gorghetto:2021fsn}.

With $N>1$, each string connects to several walls, forming a stable string-wall system.  This system would come to dominate the energy density of the Universe, leading to an unacceptable cosmology unless it disappears early enough~\cite{Zeldovich:1974uw}.
A ``bias''---a small energy difference between the vacua at both sides of each wall---would
accelerate each wall towards its adjacent higher energy vacuum, driving the domain walls towards their annihilation~\cite{Zeldovich:1974uw} (see also e.g. Ref.~\cite{Gelmini:1988sf}). 
An  additional explicit breaking term in the scalar potential was thus proposed to produce this bias~\cite{Sikivie:1982qv,Chang:1998bq}.
This term leads to the existence of one true vacuum, and a bias that we parameterize as $\epsilon_b v^4$, with a dimensionless positive parameter $\epsilon_b \ll 1$.

Gravitational waves due to cosmic strings have been recently studied for NG boson models~\cite{Chang:2019mza} and $N=1$ ALP  models~\cite{Gorghetto:2021fsn}.
We focus on models with $N>1$, in which for small enough values of $\epsilon_b$, 
GWs are dominantly produced when the string-wall system annihilates.

\section{ALP models and their cosmology}
A generic parameterization for the potential $V(\phi)$ of a pseudo-NG boson model with multiple vacua, and a small bias among them to make the model cosmologically viable (see e.g.~\cite{Sikivie:1982qv,Chang:1998bq}),  includes the terms
\begin{align}
\label{eq:potential}
V(\phi) \supset~ &  \frac{\lambda}{4} (|\phi |^2- V^2)^2 + \frac{v^4}{2} \left(1- \frac{|\phi |}{V} \cos(N \theta) \right) \nonumber\\
 - & \epsilon_b  v^4 \frac{|\phi |}{V}  \cos \left(\theta - \delta\right)~,
\end{align}
where $\phi= |\phi | e^{i \theta}$, $v \ll V$, and $V\lesssim 10^{16}~\mathrm{GeV}$ due to upper bounds on the inflation scale~\cite{Hertzberg:2008wr,Aghanim:2018eyx}.   The first term is $U(1)$ invariant. It leads to the spontaneous breaking of this symmetry at a temperature $T \simeq V$. Shortly after, $|\phi |= V$, and the phase $\theta= a/V$ has different random values in different patches of the Universe, which leads to the formation of cosmic strings.  
We assume the bosons have the same temperature or average energy as visible sector particles before the spontaneous breaking, as happens in many inflationary models.
 In a short time, the Hubble expansion and string recombination lead the string system to a scaling regime, in which the population of strings  remains of $\mathcal{O}(1)$ per Hubble volume.

 \begin{figure}[t]\centering\vspace{-5em}\hspace{-1em}
\includegraphics[width=0.49\textwidth]{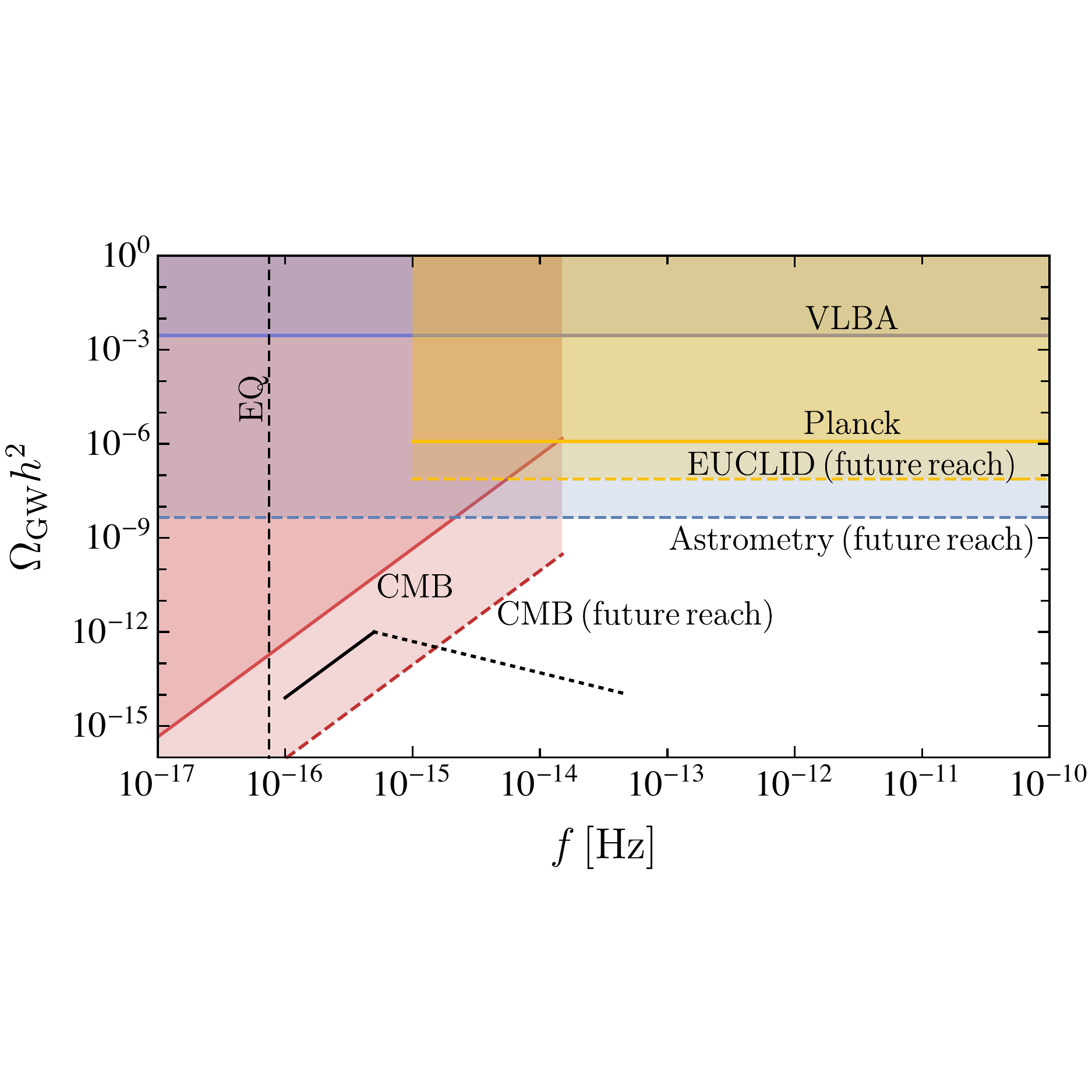}
\vspace{-5em}
\caption{Regions of $\Omega_{\rm GW}h^2$ vs. GW frequency, excluded by existing bounds (solid colored lines) or within the expected reach  (dashed colored lines) of $N_{\mathrm{eff}}$~\cite{Pagano:2015hma,Laureijs:2011gra} (yellow), astrometry~\cite{Darling:2018hmc,Arvanitaki:2019rax} (blue) and CMB~\cite{Namikawa:2019tax} (red) measurements. An example of a differential spectrum from string-wall annihilation is shown,  with $T_{\rm ann}=5$ eV and $\Omega_{\rm GW}h^2|_{\rm peak}\simeq 10^{-12}$: 
the $\sim f^3$ spectrum (solid black line) below the peak is predicted by causality while the  $\sim f^{-1}$ spectrum above the peak (dotted black line) is uncertain.  The vertical dashed line indicates the frequency of  GWs produced at matter-radiation equality.
}
\label{fig:differentialspectrum}
\end{figure}

The second term in Eq.~\eqref{eq:potential} breaks $U(1)$ into a $Z_{N}$ discrete subgroup. It produces $N$ degenerate minima with different values of $\theta$, and an ALP mass  $m_a = v^2 N/ (\sqrt{2} V)$. We assume that $a$ couplings are small enough so that temperature corrections to $m_a$ are negligible. 

At this point, the equation of motion of the field $a$ in the expanding Universe is that of a harmonic oscillator with damping term $3H \dot{a}$, where $H= (2 t)^{-1}$ is the Hubble expansion rate during the radiation dominated epoch. At a temperature $T_{w}$ when  $H(T_{w}) \simeq m_a/3$,
\begin{equation} \label{Tw}
T_{ w} \simeq \frac{5.1 \times 10^4 {\rm GeV}}{\left[g_\star(T_w)\right]^{1/4}}  \left(\frac{m_a}{\rm eV}\right)^{1/2},
\end{equation}
regions of the Universe with different values of $\theta$ evolve to different minima and become separated by domain walls of  mass per unit area $\sigma= f_\sigma v^2 V/ N$. Here $f_\sigma$ is a model dependent dimensionless parameter ($\simeq 6$ for $N=2$). Our figures assume $N=6$  and  $f_\sigma/N \simeq 1$.  
In a short time, the expansion of the Universe and energy losses drive the string-wall system into a scaling regime, in which the energy density is $\rho_w \simeq~\sigma /t$.

The third term in Eq.~\eqref{eq:potential}~\cite{Sikivie:1982qv}, assumed to be much smaller than the second one, i.e. $\epsilon_b \ll 1$, makes the vacuum closest to the arbitrary fixed phase $\delta$ the true one, and raises the others by an energy density difference, a bias, of order  $ V_{\rm bias} \simeq~\epsilon_b v^4$. We remain agnostic about the origin of this term (see e.g. Refs.~\cite{Rai:1992xw,Ringwald:2015dsf, Ferrer:2018uiu,Caputo:2019wsd}).

 The surface tension of the walls tends to rapidly straighten out curved walls to the horizon scale $H^{-1}$, and produces a pressure $p_T \simeq \sigma/t$, which decreases with time. The volume pressure $p_V \simeq V_{\rm bias}$   tends instead to accelerate the walls towards their lower energy adjacent vacuum, converting the higher energy vacuum into the lower energy one. Assuming  that when walls form $p_V \ll  p_T$ (i.e. $\epsilon_b \ll 1$), at a later time, when  $p_T \simeq p_V$, the bias drives the walls (and the strings bounding them) to annihilate within a Hubble time, when the temperature is  
\begin{equation} \label{Tann}
T_{\rm ann} \simeq \frac{0.73 \times 10^5 ~{\rm GeV}}{[g_\star(T_{\rm ann})]^{1/4}}~  \sqrt{\frac{\epsilon_b ~ m_a}{f_\sigma ~ {\rm eV}}}~.
\end{equation}
At this point the energy stored in the string-wall system goes entirely into GWs and  non-relativistic or mildly relativistic ALPs (since the wall thickness is $\simeq m_a^{-1}$)~\cite{Chang:1998tb}. 
 
\section{ Present GW energy density}

The quadrupole formula for the power emitted in GWs $P\simeq G\dddot{Q}_{ij}\dddot{Q}_{ij}$ is used to estimate the GW energy produced by the string-wall system~\cite{Maggiore:1900zz}.
 In the scaling regime the linear size of large walls is $\simeq~t$, thus their quadrupole moment as function of the energy in the walls $E_{w} \simeq \sigma t^2$ is $Q_{ij} \simeq~E_{w} t^2$.  Thus $\dddot{Q}_{ij} \simeq \sigma t$, and the power emitted in GWs is  $P \simeq G \sigma^2 t^2$.  The energy density $\Delta \rho_{\rm GW}$ emitted in a time interval $\Delta t$ is then  
 $\Delta \rho_{\rm GW} (t) \simeq G \sigma^2 (\Delta t/ t) $.
The resulting emitted energy density in a Hubble time  $\Delta t = t$ is $\simeq G \sigma^2$, independently of
the emission time, and for later emission it is less red-shifted.
Therefore, the largest contribution to the present GW energy density spectrum, the peak amplitude, corresponds to the time of wall annihilation (a similar calculation can be found e.g. in Ref.~\cite{Gelmini:2020bqg}),
\begin{equation} \label{eq:OmegaGW-walls}
\Omega_{\rm GW}h^2|_{\rm peak}\simeq \frac{1.2 \times 10^{-79} \epsilon_{gw}~ g_\star(T_{\rm ann})}{\epsilon_b^{2}~\left[g_{s \star}(T_{\rm ann}) \right]^{4/3}}
\left(\frac{f_\sigma V}{N {\text{GeV}}}\right)^4
\end{equation}
($g_\star$ and $g_{s \star}$ are the energy and entropy density numbers of degrees of freedom), see also Ref.~\cite{Hiramatsu:2012sc}. 
We include  in Eq.~\eqref{eq:OmegaGW-walls} a dimensionless factor $\epsilon_{gw}$ found in numerical simulations (see Fig.~8 of Ref.~\cite{Hiramatsu:2012sc}) and conservatively take $\epsilon_{gw}=10$.

Notice that $\Delta \rho_{\rm GW} (t)$ above defines also the maximum of the GW energy density spectrum at time $t$ as a function of the wave-number at present $k$ (which, when  defining $R_0=1$, coincides with the comoving wave-number) or of the frequency $f=k/(2\pi)$, which is defined as
$\Omega_{\rm GW}h^2 (k, t) = \left({h^2}/{\rho_c(t)}\right)
 \left({d \rho_{\rm GW} (t)}/{d \ln k}\right)$, i.e. ${d\rho_{\rm GW}(t)}/{d\ln(k)}\simeq G \sigma^2$
(see e.g. Ref.~\cite{Gelmini:2020bqg}).  Thus the peak amplitude of this GW spectrum at present, for $t= t_0$, coincides with the total amplitude in Eq.~\eqref{eq:OmegaGW-walls}.

The peak GW density is emitted at annihilation with frequency  $\simeq H(T_{\rm ann})$, which is redshifted to
\begin{equation} \label{f-peak}
f_{\rm peak}\simeq  0.76 \times 10^{-7} \text{Hz} ~\frac{T_{\rm ann}}{\rm GeV}~ \frac{\left[g_\star(T_{\rm ann})\right]^{1/2}}{\left[g_{s \star}(T_{\rm ann}) \right]^{1/3}}
\end{equation}
at present. The limit $T_{\rm ann} >$ 5 eV (safely above matter-radiation equality) thus implies $f_{\rm peak} > 5\times 10^{-16}$ Hz.  

The GW spectrum emitted by  cosmic walls for $N>1$ computed numerically is shown in Fig.~6 of Ref.~\cite{Hiramatsu:2012sc}.
It has a peak at $f_{\rm peak} \simeq R(t_f)\, m_a$ and a bump at $f\simeq R(t_f) H(t_f)$, where $t_f$ is the latest time in their simulation.  Frequencies $f < f_{\rm peak}$ correspond to
super-horizon wavelengths at annihilation, so causality requires a $\sim f^3$ dependence~\cite{Caprini:2009fx} for wavelengths that enter into the horizon during radiation domination, see e.g.~\cite{Barenboim:2016mjm,Cai:2019cdl, Hook:2020phx}.
 For $f > f_{\rm peak}$ the spectrum depends instead on the particular production model.
Reference~\cite{Hiramatsu:2012sc} finds a  $1/f$ dependence, 
although the approximate slope and height of the bump depend on $N$.

An example of the approximate spectrum is shown in Fig.~\ref{fig:differentialspectrum}, together with several present bounds and projected reaches of GW detection in the near future. For $f> 10^{-14}$ Hz, the most important bounds come from the Very Long Baseline Array (VLBA) astrometric catalog~\cite{Darling:2018hmc} (since GWs would produce an apparent distortion of the position of background sources) and  from the effective number of neutrino species $N_{\mathrm{eff}}$ during CMB emission~\cite{Pagano:2015hma} (since GWs are a radiation component).
EUCLID will improve this latter limit
by one order of magnitude~\cite{Laureijs:2011gra}, and astrometry could reach $\Omega_{\mathrm{GW}}\simeq 10^{-8}$~\cite{Arvanitaki:2019rax}. At lower frequencies, GWs are constrained by CMB polarization data~\cite{Kamionkowski:1999qc,Smith:2005mm,Clarke:2020bil,Lasky:2015lej,Campeti:2020xwn}. 
The present bounds from Planck temperature~\cite{Aghanim:2018eyx} and BICEP/Keck Array polarization~\cite{Ade:2018gkx} data sets could be improved by planned experiments such as LiteBIRD~\cite{Matsumura:2013aja}, PICO~\cite{Hanany:2019lle}, and CORE~\cite{Delabrouille:2017rct}.
We show the CMB constraints and projections of Ref.~\cite{Namikawa:2019tax} for monochromatic GWs, which may be closer to the peaked spectrum of our model than the usually assumed power-law spectrum. 

  The spectrum of GWs emitted during radiation domination by strings before walls are formed, computed in Refs.~\cite{Chang:2019mza,Gouttenoire:2019kij,Gorghetto:2021fsn}, can be approximated with the simple expression
\begin{equation} \label{OmegaGW-strings}
\Omega^{\rm st}_{\rm GW} h^2 \simeq 2 \times 10^{-15} \left(\frac{10^{-12} ~ {\rm Hz}}{f} \right)^{1/8} \left( \frac{V}{10^{14}~ {\rm GeV}} \right)^4.
\end{equation}
This spectrum, very different from the peaked spectrum produced by the string-wall network, does not extend to $f< 10^{-12}$~Hz. In fact,
the Ly-$\alpha$ limit on ALP DM $m_a> 2 \times 10^{-20}$ eV~\cite{Rogers:2020ltq} imposes a limit $T_w> 5.3$ keV (see Eq.~\eqref{Tw}) which,  replacing $T_{\rm ann}$ by $T_w$ in Eq.~\eqref{f-peak}, implies  $f > 4.7 \times 10^{-11}~\rm{Hz}$. 

The spectrum cuts off at higher $f$ for larger $m_a$ (see Fig.~4 of Ref.~\cite{Gorghetto:2021fsn}). Therefore, in our model the only source of GWs with $f < 10^{-12}~\rm{Hz}$ is the string-wall system. As clearly shown in Ref.~\cite{Gorghetto:2021fsn} for $N=1$ only a spontaneuos breaking scale   $V\gtrsim10^{14}\,\rm GeV$ and $m_a\lesssim 10^{-17}\,\rm eV$ can give an observable signal.
Thus, for lower breaking scales and heavier ALPs, the only hope to detect GWs associated to ALP production is within the scenario we consider here, with $N>1$. 

\section{Present ALP energy density}
Specifying to our model the analytic derivations in the literature (see e.g. Refs.~\cite{Hiramatsu:2010yu, Hiramatsu:2012sc,Gouttenoire:2019kij,Gorghetto:2020qws} and references therein) we obtain the different components of the present ALP density.

\begin{figure}\centering\vspace{-4
em}\hspace{-1em}
\includegraphics[width=0.48\textwidth]{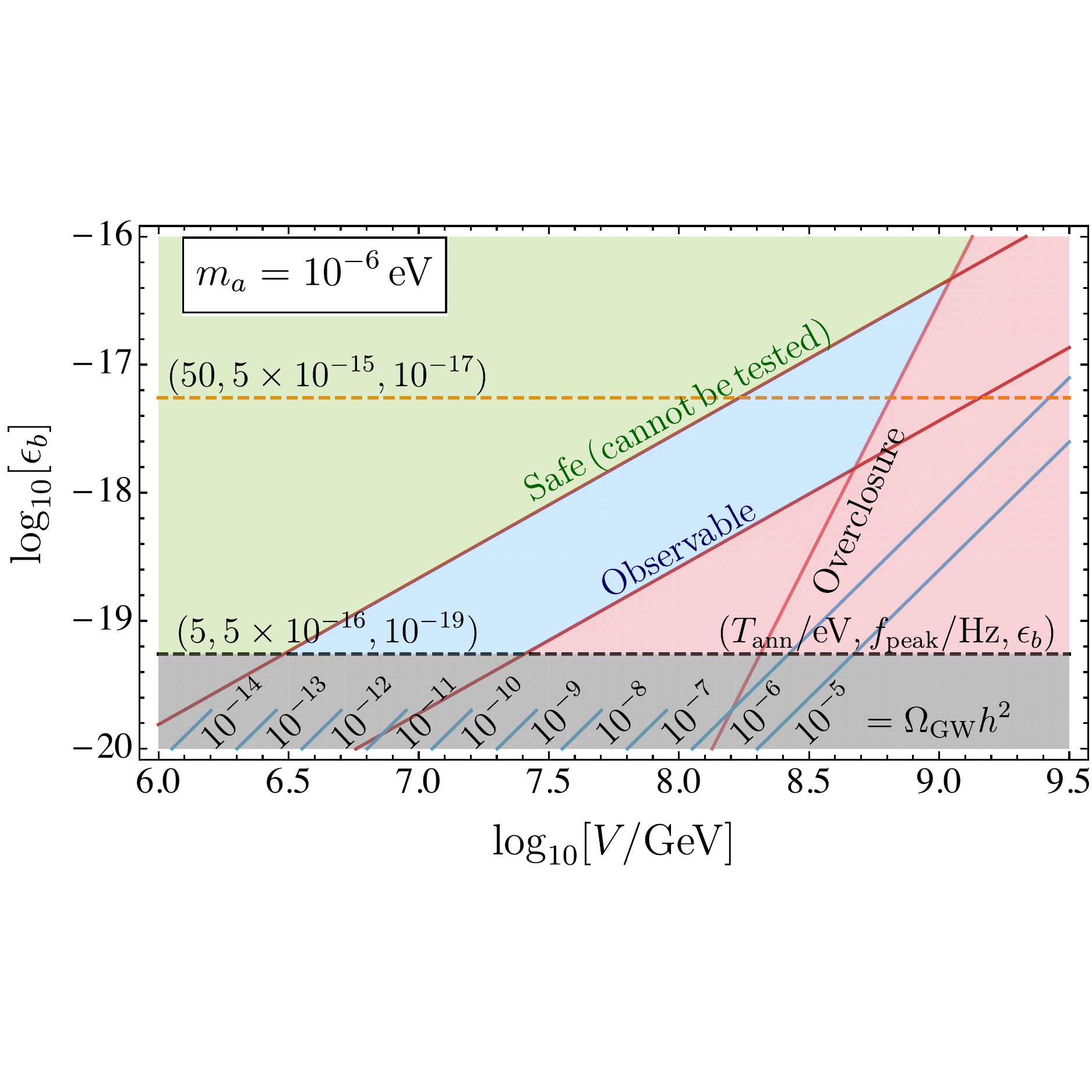}
\vspace{-5em}
\caption{Regions of interest in $\{V, \epsilon_b\}$ space for $m_a=10^{-6}$ eV. Red regions are excluded by either an ALP density larger than that of DM or current CMB limits on GWs in Fig.~\ref{fig:differentialspectrum}. The grey region corresponds to $T_{\rm ann}<5$ eV.
The blue region will be explored in the near future by CMB probes and astrometry. The green region is allowed but not testable. }
\label{fig:fixedmass6new}
\end{figure}

\begin{figure*}
\centering\vspace{-8em}
\includegraphics[width=0.65\textwidth]{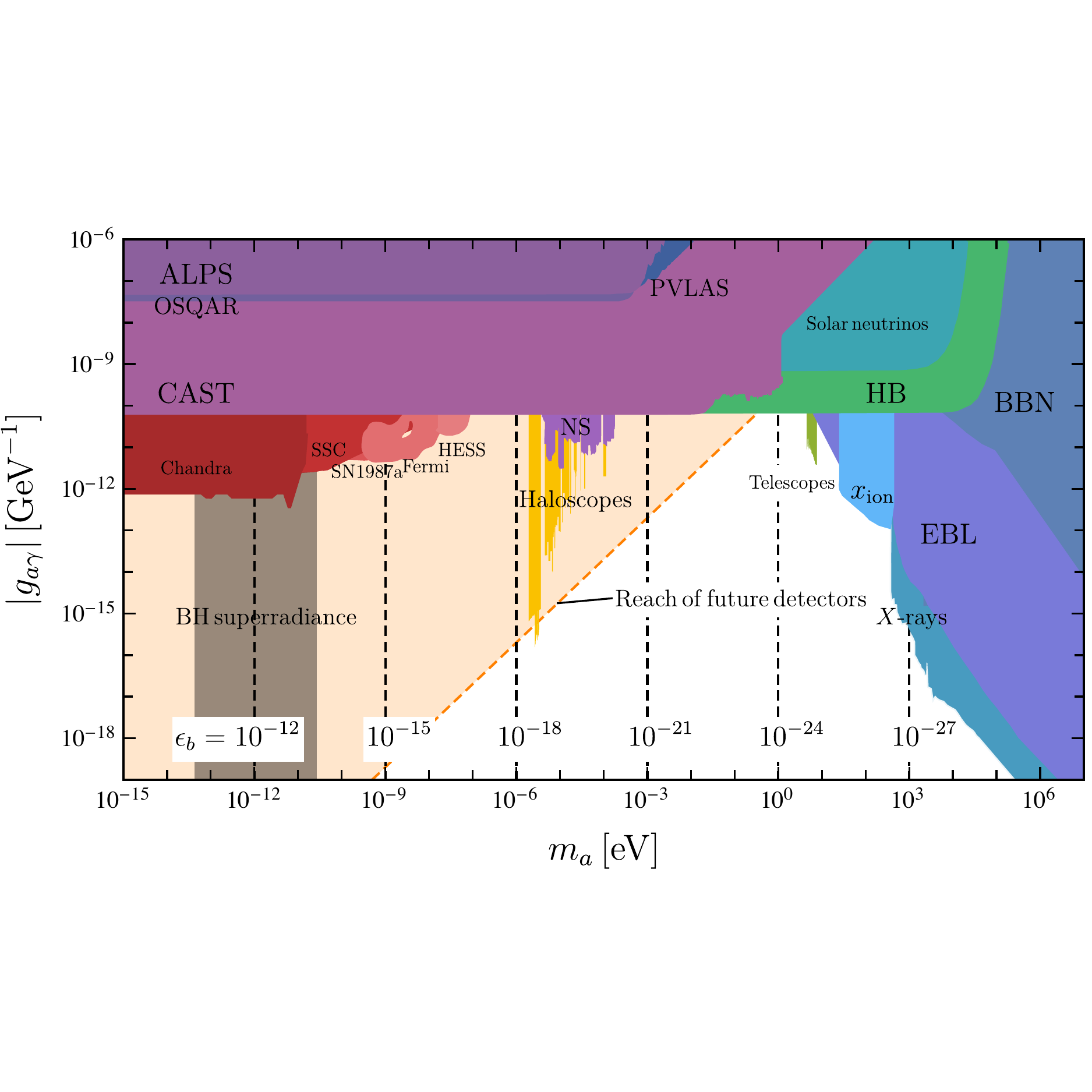}\vspace{-6em}
\caption{Parameter space $\{m_a,g_{a\gamma}\}$ for ALPs with coupling $g_{a\gamma}aF\Tilde{F}$ 
to two photons  and current bounds from laboratory~\cite{Ballou:2015cka,DellaValle:2015xxa,Ehret:2010mh}, stellar~\cite{Ayala:2014pea,Vinyoles:2015aba}, other astrophysical~\cite{Payez:2014xsa,2020ApJ...890...59R,Dessert:2020lil,Abramowski:2013oea,Foster:2020pgt,TheFermi-LAT:2016zue} and cosmological~\cite{Cadamuro:2011fd,Regis:2020fhw,Grin:2006aw} measurements, helioscopes~\cite{Andriamonje:2007ew,Anastassopoulos:2017ftl}, and direct DM detection~\cite{Asztalos:2009yp,Du:2018uak,Braine:2019fqb,Boutan:2018uoc,Lee:2020cfj,Zhong:2018rsr,Backes:2020ajv,PhysRevD.42.1297,McAllister:2017lkb,Alesini:2019ajt,Schutte-Engel:2021bqm} (see~\href{https://cajohare.github.io/AxionLimits/}{this https URL}). For each $m_a$, there is a range  of about two  orders of magnitude centered at the given $\epsilon_b$ in which GWs can  be detected. This range is independent of $g_{a\gamma}$, as are the Black Hole superradiance limits. The latter exclude the brown band
~\cite{Stott:2020gjj} (and e.g. Refs.~\cite{Arvanitaki:2009fg,Arvanitaki:2010sy,Brito:2015oca,Ikeda:2018nhb,Baryakhtar:2020gao,Blas:2020nbs,Fukuda:2019ewf}). The light orange region will be probed by future experiments, e.g.~\cite{Ouellet:2018beu,Shilon:2012te,Marsh:2018dlj,Lawson:2019brd,TheMADMAXWorkingGroup:2016hpc,Baryakhtar:2018doz,Stern:2016bbw,Alesini:2017ifp}.}
\label{fig:totalbounds}
\end{figure*}

The ALPs are produced by the string-wall system mostly at annihilation, with average energy $\simeq \sqrt{2} m_a$, 
\begin{equation} \label{Omega-a-walls}
\Omega_a h^2 \simeq \frac{2.4 \times 10^{-24}}{\epsilon_b^{1/2}}
\left(\frac{f_\sigma^{3/4} V }{N {\rm GeV}}\right)^2 \left(\frac{m_a}{\rm eV}\right)^{1/2}
\frac{[g_\star(T_{\rm ann})]^{3/4}}{g_{s \star}(T_{\rm ann})} ~.
\end{equation}
Comparing this result with Eq.~(19) of Ref.~\cite{Gorghetto:2021fsn}, we find that for
  $\epsilon_b \lesssim 2 \times 10^{-9}$ the string-wall ALP production dominates over that of strings, which emit ALPs continuously until walls form.   The component of the ALP density due to the initial misalignment of the ALP field is always subdominant.
The single contributions to the axion population due to misalignment and string decay are affected by large uncertainties (see e.g.~\cite{Gorghetto:2018myk,Klaer:2017qhr}). However, for small enough values of $\epsilon_b$ (like those considered in our figures), wall annihilation is in any event the dominant production mechanism.

Combining Eqs.~\eqref{Tann}, \eqref{eq:OmegaGW-walls}, \eqref{f-peak} and  \eqref{Omega-a-walls}, the overdensity limit $\Omega_a h^2 < \Omega_{\rm DM} h^2$ 
implies (neglecting degrees of freedom)
\begin{equation} \label{window}
\frac{\Omega_{\rm GW}h^2|_{\rm peak}}{10^{-15}} \left(\frac{f_{\rm peak}}{10^{-9} {\rm Hz}}\right)^2 < 10^{-4}~,
\end{equation}
which shows that our allowed  window is at frequencies below the $10^{-9}-10^{3}~\mathrm{Hz}$ range observable in direct GW detection for  $\Omega_{\rm GW}h^2>{10^{-15}}$. For example, for the  future reach of astrometry, $\Omega_{\rm GW}h^2 \simeq 10^{-9}$, Eq.~\eqref{window} implies $f_{\rm peak} < 10^{-14}~\mathrm{Hz}$.

\section{ GW observability}
The region of the  $\{\epsilon_b,V\}$ space which can be explored by forthcoming measurements of low frequency GWs depends on $m_a$. It is shown in blue in Fig.~\ref{fig:fixedmass6new} for $m_a= 10^{-6}$~eV. The GWs are observable
for $5 \times 10^{-16}~\mathrm{Hz}<f_{\rm peak}<1 \times 10^{-14}~\mathrm{Hz}$, i.e. for $5~\mathrm{eV} <T_{\rm ann}< 10^2~\mathrm{eV}$.
Note from  Eqs.~\eqref{Tann} and \eqref{f-peak} that $f_{\rm{peak}} \sim \epsilon_b^{1/2}$, as shown in Fig.~\ref{fig:fixedmass6new}. 
The red region is excluded either by ALPs
overclosing the Universe
or by the current CMB limits shown in Fig.~\ref{fig:differentialspectrum},  the grey region is excluded because $T_{\rm ann} \lesssim 5~\mathrm{eV}$, and the green region is allowed but the GW energy density is too small to be detected in the near future. ALPs constitute the whole of the DM on the ``Overclosure" line.

Combining Eqs.~\eqref{Tann}, \eqref{eq:OmegaGW-walls}, \eqref{f-peak} and \eqref{Omega-a-walls} one sees that the observable region shifts  
as $V \sim m_a^{-1/2}$ and $\epsilon_b \sim m_a^{-1}$. 
  As $m_a$ increases, the lowest $V$ value of the window decreases as $V=  10^{6.5} {\rm GeV} (10^{-6} {\rm eV}/m_a)^{1/2}$. Considering the hierarchy of the terms in Eq.~\eqref{eq:potential} we require $v< 10^{-2} V$, i.e. $m_a < 10^{-4} N V$. For $N=6$,  this limit restricts the observable window to $ V>2.5$ GeV and $m_a < 1.5$~MeV.
  
  The scaling of the characteristic $\epsilon_b$
  of the observable window,   
  $\epsilon_b= 10^{-18} (10^{-6} {\rm eV}/ m_a)$, shows that ALPS are dominantly produced by walls for $m_a> 5 \times 10^{-16}~\mathrm{eV} $ (for which $ \epsilon_b < 2 \times 10^{-9}$). Thus, the observable region in Fig.~\ref{fig:fixedmass6new} just translates with the same shape for $m_a\gtrsim 10^{-16}~\mathrm{eV}$. 
 We do not consider 
 lighter ALPs to avoid Black Hole superradiance limits, which also reject 
$3.8 \times 10^{ - 14}\,\mathrm{eV}<m_a<3.4 \times 10^{-11}\,\mathrm{eV}$~\cite{Stott:2020gjj}.

Figure~\ref{fig:totalbounds} shows ALP-photon-photon coupling limits and the characteristic $\epsilon_b$ for observable GWs as functions of~$m_a$. 
The $\epsilon_b$ range of observable GWs centered at the value shown is about two orders of magnitude wide.
As this range depends only on $m_a$, it applies to all ALP couplings (e.g. Refs.~\cite{Sikivie:2020zpn,Irastorza:2018dyq,OHare:2020wah}) or ``darker'' ALPs~\cite{Kaneta:2016wvf,Kalashev:2018bra,Arias:2020tzl,deNiverville:2020qoo}.

If future laboratory searches have a signal compatible with a
QCD axion, the detection of GWs with  the spectrum we described would challenge the attribution of this signal to a QCD axion, since GWs from QCD axion string-wall networks are not detectable~\cite{Hiramatsu:2012sc}.

\section{ Conclusions}
We have presented a novel window onto ALP models which takes advantage of the fast progress expected in GW detection, resulting from a so far overlooked mechanism of GW production in ALP models.
If the ALP potential has several minima, a bias between them is needed to drive the ensuing string-wall system to annihilate early enough to avoid cosmological problems. For the QCD axion, GWs generated by this mechanism are unobservable, but for other ALPs it could produce GWs is a novel frequency range not previously identified for ALP models. 
We have found that, if walls annihilate at 5~eV $\lesssim T_{\mathrm{ann}}\lesssim 10^{2}$ eV, GWs can potentially be detected by future CMB probes and astrometry measurements for ALPs with mass from $10^{-16}$ to $10^{6}~\mathrm{eV}$, which could constitute all of the DM.

\smallskip
 \section{Acknowledgments}
{The work of GG and EV was supported in part by the U.S. Department of Energy (DOE) Grant No. DE-SC0009937.}
  \appendix

\bibliographystyle{bibi}
\bibliography{bibliography}

\providecommand{\href}[2]{#2}\begingroup\raggedright\begin{thebibliography}{100}

\bibitem{Vilenkin:2000jqa}
A.~Vilenkin and E.~P.~S. Shellard, \emph{{Cosmic Strings and Other Topological
  Defects}}. Cambridge University Press, 7, 2000.

\bibitem{Maggiore:1900zz}
M.~Maggiore, \emph{{Gravitational Waves. Vol. 1: Theory and Experiments}},
  Oxford Master Series in Physics. Oxford University Press, 2007.

\bibitem{Maggiore:2018sht}
M.~Maggiore, \emph{{Gravitational Waves. Vol. 2: Astrophysics and Cosmology}}.
  Oxford University Press, 3, 2018.

\bibitem{Sathyaprakash:2009xs}
B.~S. Sathyaprakash and B.~F. Schutz, \emph{{Physics, Astrophysics and
  Cosmology with Gravitational Waves}},
  \href{https://doi.org/10.12942/lrr-2009-2}{\emph{Living Rev. Rel.} {\bfseries
  12} (2009) 2} [\href{https://arxiv.org/abs/0903.0338}{{\ttfamily
  0903.0338}}].

\bibitem{Barack:2018yly}
L.~Barack et~al., \emph{{Black holes, gravitational waves and fundamental
  physics: a roadmap}},
  \href{https://doi.org/10.1088/1361-6382/ab0587}{\emph{Class. Quant. Grav.}
  {\bfseries 36} (2019) 143001}
  [\href{https://arxiv.org/abs/1806.05195}{{\ttfamily 1806.05195}}].

\bibitem{Peccei:1977hh}
R.~D. Peccei and H.~R. Quinn, \emph{{CP Conservation in the Presence of
  Instantons}}, \href{https://doi.org/10.1103/PhysRevLett.38.1440}{\emph{Phys.
  Rev. Lett.} {\bfseries 38} (1977) 1440}.

\bibitem{Weinberg:1977ma}
S.~Weinberg, \emph{{A New Light Boson?}},
  \href{https://doi.org/10.1103/PhysRevLett.40.223}{\emph{Phys. Rev. Lett.}
  {\bfseries 40} (1978) 223}.

\bibitem{Wilczek:1977pj}
F.~Wilczek, \emph{{Problem of Strong $P$ and $T$ Invariance in the Presence of
  Instantons}}, \href{https://doi.org/10.1103/PhysRevLett.40.279}{\emph{Phys.
  Rev. Lett.} {\bfseries 40} (1978) 279}.

\bibitem{Kim:1979if}
J.~E. Kim, \emph{{Weak Interaction Singlet and Strong CP Invariance}},
  \href{https://doi.org/10.1103/PhysRevLett.43.103}{\emph{Phys. Rev. Lett.}
  {\bfseries 43} (1979) 103}.

\bibitem{Shifman:1979if}
M.~A. Shifman, A.~I. Vainshtein and V.~I. Zakharov, \emph{{Can Confinement
  Ensure Natural CP Invariance of Strong Interactions?}},
  \href{https://doi.org/10.1016/0550-3213(80)90209-6}{\emph{Nucl. Phys. B}
  {\bfseries 166} (1980) 493}.

\bibitem{Dine:1981rt}
M.~Dine, W.~Fischler and M.~Srednicki, \emph{{A Simple Solution to the Strong
  CP Problem with a Harmless Axion}},
  \href{https://doi.org/10.1016/0370-2693(81)90590-6}{\emph{Phys. Lett. B}
  {\bfseries 104} (1981) 199}.

\bibitem{Zhitnitsky:1980tq}
A.~R. Zhitnitsky, \emph{{On Possible Suppression of the Axion Hadron
  Interactions. (In Russian)}}, {\emph{Sov. J. Nucl. Phys.} {\bfseries 31}
  (1980) 260}.

\bibitem{Svrcek:2006yi}
P.~Svrcek and E.~Witten, \emph{{Axions In String Theory}},
  \href{https://doi.org/10.1088/1126-6708/2006/06/051}{\emph{JHEP} {\bfseries
  06} (2006) 051} [\href{https://arxiv.org/abs/hep-th/0605206}{{\ttfamily
  hep-th/0605206}}].

\bibitem{Arvanitaki:2009fg}
A.~Arvanitaki, S.~Dimopoulos, S.~Dubovsky, N.~Kaloper and J.~March-Russell,
  \emph{{String Axiverse}},
  \href{https://doi.org/10.1103/PhysRevD.81.123530}{\emph{Phys. Rev. D}
  {\bfseries 81} (2010) 123530}
  [\href{https://arxiv.org/abs/0905.4720}{{\ttfamily 0905.4720}}].

\bibitem{Acharya:2010zx}
B.~S. Acharya, K.~Bobkov and P.~Kumar, \emph{{An M Theory Solution to the
  Strong CP Problem and Constraints on the Axiverse}},
  \href{https://doi.org/10.1007/JHEP11(2010)105}{\emph{JHEP} {\bfseries 11}
  (2010) 105} [\href{https://arxiv.org/abs/1004.5138}{{\ttfamily 1004.5138}}].

\bibitem{Dine:2010cr}
M.~Dine, G.~Festuccia, J.~Kehayias and W.~Wu, \emph{{Axions in the Landscape
  and String Theory}},
  \href{https://doi.org/10.1007/JHEP01(2011)012}{\emph{JHEP} {\bfseries 01}
  (2011) 012} [\href{https://arxiv.org/abs/1010.4803}{{\ttfamily 1010.4803}}].

\bibitem{Jaeckel:2010ni}
J.~Jaeckel and A.~Ringwald, \emph{{The Low-Energy Frontier of Particle
  Physics}},
  \href{https://doi.org/10.1146/annurev.nucl.012809.104433}{\emph{Ann. Rev.
  Nucl. Part. Sci.} {\bfseries 60} (2010) 405}
  [\href{https://arxiv.org/abs/1002.0329}{{\ttfamily 1002.0329}}].

\bibitem{Vilenkin:1984ib}
A.~Vilenkin, \emph{{Cosmic Strings and Domain Walls}},
  \href{https://doi.org/10.1016/0370-1573(85)90033-X}{\emph{Phys. Rept.}
  {\bfseries 121} (1985) 263}.

\bibitem{Gorghetto:2021fsn}
M.~Gorghetto, E.~Hardy and H.~Nicolaescu, \emph{{Observing Invisible Axions
  with Gravitational Waves}},
  \href{https://arxiv.org/abs/2101.11007}{{\ttfamily 2101.11007}}.

\bibitem{Zeldovich:1974uw}
Y.~Zeldovich, I.~Kobzarev and L.~Okun, \emph{{Cosmological Consequences of the
  Spontaneous Breakdown of Discrete Symmetry}}, {\emph{Zh. Eksp. Teor. Fiz.}
  {\bfseries 67} (1974) 3}.

\bibitem{Gelmini:1988sf}
G.~B. Gelmini, M.~Gleiser and E.~W. Kolb, \emph{{Cosmology of Biased Discrete
  Symmetry Breaking}},
  \href{https://doi.org/10.1103/PhysRevD.39.1558}{\emph{Phys. Rev. D}
  {\bfseries 39} (1989) 1558}.

\bibitem{Sikivie:1982qv}
P.~Sikivie, \emph{{Of Axions, Domain Walls and the Early Universe}},
  \href{https://doi.org/10.1103/PhysRevLett.48.1156}{\emph{Phys. Rev. Lett.}
  {\bfseries 48} (1982) 1156}.

\bibitem{Chang:1998bq}
S.~Chang, C.~Hagmann and P.~Sikivie, \emph{{Axions from wall decay}},
  \href{https://doi.org/10.1016/S0920-5632(98)00510-6}{\emph{Nucl. Phys. B
  Proc. Suppl.} {\bfseries 72} (1999) 99}
  [\href{https://arxiv.org/abs/hep-ph/9808302}{{\ttfamily hep-ph/9808302}}].

\bibitem{Chang:2019mza}
C.-F. Chang and Y.~Cui, \emph{{Stochastic Gravitational Wave Background from
  Global Cosmic Strings}},
  \href{https://doi.org/10.1016/j.dark.2020.100604}{\emph{Phys. Dark Univ.}
  {\bfseries 29} (2020) 100604}
  [\href{https://arxiv.org/abs/1910.04781}{{\ttfamily 1910.04781}}].

\bibitem{Hertzberg:2008wr}
M.~P. Hertzberg, M.~Tegmark and F.~Wilczek, \emph{{Axion Cosmology and the
  Energy Scale of Inflation}},
  \href{https://doi.org/10.1103/PhysRevD.78.083507}{\emph{Phys. Rev. D}
  {\bfseries 78} (2008) 083507}
  [\href{https://arxiv.org/abs/0807.1726}{{\ttfamily 0807.1726}}].

\bibitem{Aghanim:2018eyx}
{\scshape Planck} Collaboration, N.~Aghanim et~al., \emph{{Planck 2018 results.
  VI. Cosmological parameters}},
  \href{https://doi.org/10.1051/0004-6361/201833910}{\emph{Astron. Astrophys.}
  {\bfseries 641} (2020) A6}
  [\href{https://arxiv.org/abs/1807.06209}{{\ttfamily 1807.06209}}].

\bibitem{Pagano:2015hma}
L.~Pagano, L.~Salvati and A.~Melchiorri, \emph{{New constraints on primordial
  gravitational waves from Planck 2015}},
  \href{https://doi.org/10.1016/j.physletb.2016.07.078}{\emph{Phys. Lett. B}
  {\bfseries 760} (2016) 823}
  [\href{https://arxiv.org/abs/1508.02393}{{\ttfamily 1508.02393}}].

\bibitem{Laureijs:2011gra}
{\scshape EUCLID} Collaboration, R.~Laureijs et~al., \emph{{Euclid Definition
  Study Report}},  \href{https://arxiv.org/abs/1110.3193}{{\ttfamily
  1110.3193}}.

\bibitem{Darling:2018hmc}
J.~Darling, A.~E. Truebenbach and J.~Paine, \emph{{Astrometric Limits on the
  Stochastic Gravitational Wave Background}},
  \href{https://doi.org/10.3847/1538-4357/aac772}{\emph{Astrophys. J.}
  {\bfseries 861} (2018) 113}
  [\href{https://arxiv.org/abs/1804.06986}{{\ttfamily 1804.06986}}].

\bibitem{Arvanitaki:2019rax}
A.~Arvanitaki, S.~Dimopoulos, M.~Galanis, L.~Lehner, J.~O. Thompson and
  K.~Van~Tilburg, \emph{{Large-misalignment mechanism for the formation of
  compact axion structures: Signatures from the QCD axion to fuzzy dark
  matter}}, \href{https://doi.org/10.1103/PhysRevD.101.083014}{\emph{Phys. Rev.
  D} {\bfseries 101} (2020) 083014}
  [\href{https://arxiv.org/abs/1909.11665}{{\ttfamily 1909.11665}}].

\bibitem{Namikawa:2019tax}
T.~Namikawa, S.~Saga, D.~Yamauchi and A.~Taruya, \emph{{CMB Constraints on the
  Stochastic Gravitational-Wave Background at Mpc scales}},
  \href{https://doi.org/10.1103/PhysRevD.100.021303}{\emph{Phys. Rev. D}
  {\bfseries 100} (2019) 021303}
  [\href{https://arxiv.org/abs/1904.02115}{{\ttfamily 1904.02115}}].

\bibitem{Rai:1992xw}
B.~Rai and G.~Senjanovic, \emph{{Gravity and domain wall problem}},
  \href{https://doi.org/10.1103/PhysRevD.49.2729}{\emph{Phys. Rev. D}
  {\bfseries 49} (1994) 2729}
  [\href{https://arxiv.org/abs/hep-ph/9301240}{{\ttfamily hep-ph/9301240}}].

\bibitem{Ringwald:2015dsf}
A.~Ringwald and K.~Saikawa, \emph{{Axion dark matter in the post-inflationary
  Peccei-Quinn symmetry breaking scenario}},
  \href{https://doi.org/10.1103/PhysRevD.93.085031}{\emph{Phys. Rev. D}
  {\bfseries 93} (2016) 085031}
  [\href{https://arxiv.org/abs/1512.06436}{{\ttfamily 1512.06436}}]. [Addendum:
  Phys.Rev.D 94, 049908 (2016)].

\bibitem{Ferrer:2018uiu}
F.~Ferrer, E.~Masso, G.~Panico, O.~Pujolas and F.~Rompineve, \emph{{Primordial
  Black Holes from the QCD axion}},
  \href{https://doi.org/10.1103/PhysRevLett.122.101301}{\emph{Phys. Rev. Lett.}
  {\bfseries 122} (2019) 101301}
  [\href{https://arxiv.org/abs/1807.01707}{{\ttfamily 1807.01707}}].

\bibitem{Caputo:2019wsd}
A.~Caputo and M.~Reig, \emph{{Cosmic implications of a low-scale solution to
  the axion domain wall problem}},
  \href{https://doi.org/10.1103/PhysRevD.100.063530}{\emph{Phys. Rev. D}
  {\bfseries 100} (2019) 063530}
  [\href{https://arxiv.org/abs/1905.13116}{{\ttfamily 1905.13116}}].

\bibitem{Chang:1998tb}
S.~Chang, C.~Hagmann and P.~Sikivie, \emph{{Studies of the motion and decay of
  axion walls bounded by strings}},
  \href{https://doi.org/10.1103/PhysRevD.59.023505}{\emph{Phys. Rev. D}
  {\bfseries 59} (1999) 023505}
  [\href{https://arxiv.org/abs/hep-ph/9807374}{{\ttfamily hep-ph/9807374}}].

\bibitem{Gelmini:2020bqg}
G.~B. Gelmini, S.~Pascoli, E.~Vitagliano and Y.-L. Zhou, \emph{{Gravitational
  wave signatures from discrete flavor symmetries}},
  \href{https://doi.org/10.1088/1475-7516/2021/02/032}{\emph{JCAP} {\bfseries
  02} (2021) 032} [\href{https://arxiv.org/abs/2009.01903}{{\ttfamily
  2009.01903}}].

\bibitem{Hiramatsu:2012sc}
T.~Hiramatsu, M.~Kawasaki, K.~Saikawa and T.~Sekiguchi, \emph{{Axion cosmology
  with long-lived domain walls}},
  \href{https://doi.org/10.1088/1475-7516/2013/01/001}{\emph{JCAP} {\bfseries
  01} (2013) 001} [\href{https://arxiv.org/abs/1207.3166}{{\ttfamily
  1207.3166}}].

\bibitem{Caprini:2009fx}
C.~Caprini, R.~Durrer, T.~Konstandin and G.~Servant, \emph{{General Properties
  of the Gravitational Wave Spectrum from Phase Transitions}},
  \href{https://doi.org/10.1103/PhysRevD.79.083519}{\emph{Phys. Rev. D}
  {\bfseries 79} (2009) 083519}
  [\href{https://arxiv.org/abs/0901.1661}{{\ttfamily 0901.1661}}].

\bibitem{Barenboim:2016mjm}
G.~Barenboim and W.-I. Park, \emph{{Gravitational waves from first order phase
  transitions as a probe of an early matter domination era and its inverse
  problem}}, \href{https://doi.org/10.1016/j.physletb.2016.06.009}{\emph{Phys.
  Lett. B} {\bfseries 759} (2016) 430}
  [\href{https://arxiv.org/abs/1605.03781}{{\ttfamily 1605.03781}}].

\bibitem{Cai:2019cdl}
R.-G. Cai, S.~Pi and M.~Sasaki, \emph{{Universal infrared scaling of
  gravitational wave background spectra}},
  \href{https://doi.org/10.1103/PhysRevD.102.083528}{\emph{Phys. Rev. D}
  {\bfseries 102} (2020) 083528}
  [\href{https://arxiv.org/abs/1909.13728}{{\ttfamily 1909.13728}}].

\bibitem{Hook:2020phx}
A.~Hook, G.~Marques-Tavares and D.~Racco, \emph{{Causal gravitational waves as
  a probe of free streaming particles and the expansion of the Universe}},
  \href{https://doi.org/10.1007/JHEP02(2021)117}{\emph{JHEP} {\bfseries 02}
  (2021) 117} [\href{https://arxiv.org/abs/2010.03568}{{\ttfamily
  2010.03568}}].

\bibitem{Kamionkowski:1999qc}
M.~Kamionkowski and A.~Kosowsky, \emph{{The Cosmic microwave background and
  particle physics}},
  \href{https://doi.org/10.1146/annurev.nucl.49.1.77}{\emph{Ann. Rev. Nucl.
  Part. Sci.} {\bfseries 49} (1999) 77}
  [\href{https://arxiv.org/abs/astro-ph/9904108}{{\ttfamily
  astro-ph/9904108}}].

\bibitem{Smith:2005mm}
T.~L. Smith, M.~Kamionkowski and A.~Cooray, \emph{{Direct detection of the
  inflationary gravitational wave background}},
  \href{https://doi.org/10.1103/PhysRevD.73.023504}{\emph{Phys. Rev. D}
  {\bfseries 73} (2006) 023504}
  [\href{https://arxiv.org/abs/astro-ph/0506422}{{\ttfamily
  astro-ph/0506422}}].

\bibitem{Clarke:2020bil}
T.~J. Clarke, E.~J. Copeland and A.~Moss, \emph{{Constraints on primordial
  gravitational waves from the Cosmic Microwave Background}},
  \href{https://doi.org/10.1088/1475-7516/2020/10/002}{\emph{JCAP} {\bfseries
  10} (2020) 002} [\href{https://arxiv.org/abs/2004.11396}{{\ttfamily
  2004.11396}}].

\bibitem{Lasky:2015lej}
P.~D. Lasky et~al., \emph{{Gravitational-wave cosmology across 29 decades in
  frequency}}, \href{https://doi.org/10.1103/PhysRevX.6.011035}{\emph{Phys.
  Rev. X} {\bfseries 6} (2016) 011035}
  [\href{https://arxiv.org/abs/1511.05994}{{\ttfamily 1511.05994}}].

\bibitem{Campeti:2020xwn}
P.~Campeti, E.~Komatsu, D.~Poletti and C.~Baccigalupi, \emph{{Measuring the
  spectrum of primordial gravitational waves with CMB, PTA and Laser
  Interferometers}},
  \href{https://doi.org/10.1088/1475-7516/2021/01/012}{\emph{JCAP} {\bfseries
  01} (2021) 012} [\href{https://arxiv.org/abs/2007.04241}{{\ttfamily
  2007.04241}}].

\bibitem{Ade:2018gkx}
{\scshape BICEP2, Keck Array} Collaboration, P.~A.~R. Ade et~al., \emph{{BICEP2
  / Keck Array x: Constraints on Primordial Gravitational Waves using Planck,
  WMAP, and New BICEP2/Keck Observations through the 2015 Season}},
  \href{https://doi.org/10.1103/PhysRevLett.121.221301}{\emph{Phys. Rev. Lett.}
  {\bfseries 121} (2018) 221301}
  [\href{https://arxiv.org/abs/1810.05216}{{\ttfamily 1810.05216}}].

\bibitem{Matsumura:2013aja}
T.~Matsumura et~al., \emph{{Mission design of LiteBIRD}},
  \href{https://doi.org/10.1007/s10909-013-0996-1}{\emph{J. Low Temp. Phys.}
  {\bfseries 176} (2014) 733}
  [\href{https://arxiv.org/abs/1311.2847}{{\ttfamily 1311.2847}}].

\bibitem{Hanany:2019lle}
{\scshape NASA PICO} Collaboration, S.~Hanany et~al., \emph{{PICO: Probe of
  Inflation and Cosmic Origins}},
  \href{https://arxiv.org/abs/1902.10541}{{\ttfamily 1902.10541}}.

\bibitem{Delabrouille:2017rct}
{\scshape CORE} Collaboration, J.~Delabrouille et~al., \emph{{Exploring cosmic
  origins with CORE: Survey requirements and mission design}},
  \href{https://doi.org/10.1088/1475-7516/2018/04/014}{\emph{JCAP} {\bfseries
  04} (2018) 014} [\href{https://arxiv.org/abs/1706.04516}{{\ttfamily
  1706.04516}}].

\bibitem{Gouttenoire:2019kij}
Y.~Gouttenoire, G.~Servant and P.~Simakachorn, \emph{{Beyond the Standard
  Models with Cosmic Strings}},
  \href{https://doi.org/10.1088/1475-7516/2020/07/032}{\emph{JCAP} {\bfseries
  07} (2020) 032} [\href{https://arxiv.org/abs/1912.02569}{{\ttfamily
  1912.02569}}].

\bibitem{Rogers:2020ltq}
K.~K. Rogers and H.~V. Peiris, \emph{{Strong bound on canonical ultra-light
  axion dark matter from the Lyman-alpha forest}},
  \href{https://doi.org/10.1103/PhysRevLett.126.071302}{\emph{Phys. Rev. Lett.}
  {\bfseries 126} (2021) 071302}
  [\href{https://arxiv.org/abs/2007.12705}{{\ttfamily 2007.12705}}].

\bibitem{Hiramatsu:2010yu}
T.~Hiramatsu, M.~Kawasaki, T.~Sekiguchi, M.~Yamaguchi and J.~Yokoyama,
  \emph{{Improved estimation of radiated axions from cosmological axionic
  strings}}, \href{https://doi.org/10.1103/PhysRevD.83.123531}{\emph{Phys. Rev.
  D} {\bfseries 83} (2011) 123531}
  [\href{https://arxiv.org/abs/1012.5502}{{\ttfamily 1012.5502}}].

\bibitem{Gorghetto:2020qws}
M.~Gorghetto, E.~Hardy and G.~Villadoro, \emph{{More Axions from Strings}},
  \href{https://arxiv.org/abs/2007.04990}{{\ttfamily 2007.04990}}.

\bibitem{Ballou:2015cka}
{\scshape OSQAR} Collaboration, R.~Ballou et~al., \emph{{New exclusion limits
  on scalar and pseudoscalar axionlike particles from light shining through a
  wall}}, \href{https://doi.org/10.1103/PhysRevD.92.092002}{\emph{Phys. Rev. D}
  {\bfseries 92} (2015) 092002}
  [\href{https://arxiv.org/abs/1506.08082}{{\ttfamily 1506.08082}}].

\bibitem{DellaValle:2015xxa}
F.~Della~Valle, A.~Ejlli, U.~Gastaldi, G.~Messineo, E.~Milotti, R.~Pengo,
  G.~Ruoso and G.~Zavattini, \emph{{The PVLAS experiment: measuring vacuum
  magnetic birefringence and dichroism with a birefringent
  Fabry\textendash{}Perot cavity}},
  \href{https://doi.org/10.1140/epjc/s10052-015-3869-8}{\emph{Eur. Phys. J. C}
  {\bfseries 76} (2016) 24} [\href{https://arxiv.org/abs/1510.08052}{{\ttfamily
  1510.08052}}].

\bibitem{Ehret:2010mh}
K.~Ehret et~al., \emph{{New ALPS Results on Hidden-Sector Lightweights}},
  \href{https://doi.org/10.1016/j.physletb.2010.04.066}{\emph{Phys. Lett. B}
  {\bfseries 689} (2010) 149}
  [\href{https://arxiv.org/abs/1004.1313}{{\ttfamily 1004.1313}}].

\bibitem{Ayala:2014pea}
A.~Ayala, I.~Dom\'\i{}nguez, M.~Giannotti, A.~Mirizzi and O.~Straniero,
  \emph{{Revisiting the bound on axion-photon coupling from Globular
  Clusters}}, \href{https://doi.org/10.1103/PhysRevLett.113.191302}{\emph{Phys.
  Rev. Lett.} {\bfseries 113} (2014) 191302}
  [\href{https://arxiv.org/abs/1406.6053}{{\ttfamily 1406.6053}}].

\bibitem{Vinyoles:2015aba}
N.~Vinyoles, A.~Serenelli, F.~L. Villante, S.~Basu, J.~Redondo and J.~Isern,
  \emph{{New axion and hidden photon constraints from a solar data global
  fit}}, \href{https://doi.org/10.1088/1475-7516/2015/10/015}{\emph{JCAP}
  {\bfseries 10} (2015) 015}
  [\href{https://arxiv.org/abs/1501.01639}{{\ttfamily 1501.01639}}].

\bibitem{Payez:2014xsa}
A.~Payez, C.~Evoli, T.~Fischer, M.~Giannotti, A.~Mirizzi and A.~Ringwald,
  \emph{{Revisiting the SN1987A gamma-ray limit on ultralight axion-like
  particles}}, \href{https://doi.org/10.1088/1475-7516/2015/02/006}{\emph{JCAP}
  {\bfseries 02} (2015) 006} [\href{https://arxiv.org/abs/1410.3747}{{\ttfamily
  1410.3747}}].

\bibitem{2020ApJ...890...59R}
C.~S. {Reynolds}, M.~C.~D. {Marsh}, H.~R. {Russell}, A.~C. {Fabian},
  R.~{Smith}, F.~{Tombesi} and S.~{Veilleux}, \emph{{Astrophysical Limits on
  Very Light Axion-like Particles from Chandra Grating Spectroscopy of NGC
  1275}}, \href{https://doi.org/10.3847/1538-4357/ab6a0c}{\emph{\apj}
  {\bfseries 890} (2020) 59}
  [\href{https://arxiv.org/abs/1907.05475}{{\ttfamily 1907.05475}}].

\bibitem{Dessert:2020lil}
C.~Dessert, J.~W. Foster and B.~R. Safdi, \emph{{X-ray Searches for Axions from
  Super Star Clusters}},
  \href{https://doi.org/10.1103/PhysRevLett.125.261102}{\emph{Phys. Rev. Lett.}
  {\bfseries 125} (2020) 261102}
  [\href{https://arxiv.org/abs/2008.03305}{{\ttfamily 2008.03305}}].

\bibitem{Abramowski:2013oea}
{\scshape H.E.S.S.} Collaboration, A.~Abramowski et~al., \emph{{Constraints on
  axionlike particles with H.E.S.S. from the irregularity of the PKS 2155-304
  energy spectrum}},
  \href{https://doi.org/10.1103/PhysRevD.88.102003}{\emph{Phys. Rev. D}
  {\bfseries 88} (2013) 102003}
  [\href{https://arxiv.org/abs/1311.3148}{{\ttfamily 1311.3148}}].

\bibitem{Foster:2020pgt}
J.~W. Foster, Y.~Kahn, O.~Macias, Z.~Sun, R.~P. Eatough, V.~I. Kondratiev,
  W.~M. Peters, C.~Weniger and B.~R. Safdi, \emph{{Green Bank and Effelsberg
  Radio Telescope Searches for Axion Dark Matter Conversion in Neutron Star
  Magnetospheres}},
  \href{https://doi.org/10.1103/PhysRevLett.125.171301}{\emph{Phys. Rev. Lett.}
  {\bfseries 125} (2020) 171301}
  [\href{https://arxiv.org/abs/2004.00011}{{\ttfamily 2004.00011}}].

\bibitem{TheFermi-LAT:2016zue}
{\scshape Fermi-LAT} Collaboration, M.~Ajello et~al., \emph{{Search for
  Spectral Irregularities due to Photon\textendash{}Axionlike-Particle
  Oscillations with the Fermi Large Area Telescope}},
  \href{https://doi.org/10.1103/PhysRevLett.116.161101}{\emph{Phys. Rev. Lett.}
  {\bfseries 116} (2016) 161101}
  [\href{https://arxiv.org/abs/1603.06978}{{\ttfamily 1603.06978}}].

\bibitem{Cadamuro:2011fd}
D.~Cadamuro and J.~Redondo, \emph{{Cosmological bounds on pseudo
  Nambu-Goldstone bosons}},
  \href{https://doi.org/10.1088/1475-7516/2012/02/032}{\emph{JCAP} {\bfseries
  02} (2012) 032} [\href{https://arxiv.org/abs/1110.2895}{{\ttfamily
  1110.2895}}].

\bibitem{Regis:2020fhw}
M.~Regis, M.~Taoso, D.~Vaz, J.~Brinchmann, S.~L. Zoutendijk, N.~F. Bouch\'e and
  M.~Steinmetz, \emph{{Searching for light in the darkness: Bounds on ALP dark
  matter with the optical MUSE-faint survey}},
  \href{https://doi.org/10.1016/j.physletb.2021.136075}{\emph{Phys. Lett. B}
  {\bfseries 814} (2021) 136075}
  [\href{https://arxiv.org/abs/2009.01310}{{\ttfamily 2009.01310}}].

\bibitem{Grin:2006aw}
D.~Grin, G.~Covone, J.-P. Kneib, M.~Kamionkowski, A.~Blain and E.~Jullo,
  \emph{{A Telescope Search for Decaying Relic Axions}},
  \href{https://doi.org/10.1103/PhysRevD.75.105018}{\emph{Phys. Rev. D}
  {\bfseries 75} (2007) 105018}
  [\href{https://arxiv.org/abs/astro-ph/0611502}{{\ttfamily
  astro-ph/0611502}}].

\bibitem{Andriamonje:2007ew}
{\scshape CAST} Collaboration, S.~Andriamonje et~al., \emph{{An Improved limit
  on the axion-photon coupling from the CAST experiment}},
  \href{https://doi.org/10.1088/1475-7516/2007/04/010}{\emph{JCAP} {\bfseries
  04} (2007) 010} [\href{https://arxiv.org/abs/hep-ex/0702006}{{\ttfamily
  hep-ex/0702006}}].

\bibitem{Anastassopoulos:2017ftl}
{\scshape CAST} Collaboration, V.~Anastassopoulos et~al., \emph{{New CAST Limit
  on the Axion-Photon Interaction}},
  \href{https://doi.org/10.1038/nphys4109}{\emph{Nature Phys.} {\bfseries 13}
  (2017) 584} [\href{https://arxiv.org/abs/1705.02290}{{\ttfamily
  1705.02290}}].

\bibitem{Asztalos:2009yp}
{\scshape ADMX} Collaboration, S.~J. Asztalos et~al., \emph{{A SQUID-based
  microwave cavity search for dark-matter axions}},
  \href{https://doi.org/10.1103/PhysRevLett.104.041301}{\emph{Phys. Rev. Lett.}
  {\bfseries 104} (2010) 041301}
  [\href{https://arxiv.org/abs/0910.5914}{{\ttfamily 0910.5914}}].

\bibitem{Du:2018uak}
{\scshape ADMX} Collaboration, N.~Du et~al., \emph{{A Search for Invisible
  Axion Dark Matter with the Axion Dark Matter Experiment}},
  \href{https://doi.org/10.1103/PhysRevLett.120.151301}{\emph{Phys. Rev. Lett.}
  {\bfseries 120} (2018) 151301}
  [\href{https://arxiv.org/abs/1804.05750}{{\ttfamily 1804.05750}}].

\bibitem{Braine:2019fqb}
{\scshape ADMX} Collaboration, T.~Braine et~al., \emph{{Extended Search for the
  Invisible Axion with the Axion Dark Matter Experiment}},
  \href{https://doi.org/10.1103/PhysRevLett.124.101303}{\emph{Phys. Rev. Lett.}
  {\bfseries 124} (2020) 101303}
  [\href{https://arxiv.org/abs/1910.08638}{{\ttfamily 1910.08638}}].

\bibitem{Boutan:2018uoc}
{\scshape ADMX} Collaboration, C.~Boutan et~al., \emph{{Piezoelectrically Tuned
  Multimode Cavity Search for Axion Dark Matter}},
  \href{https://doi.org/10.1103/PhysRevLett.121.261302}{\emph{Phys. Rev. Lett.}
  {\bfseries 121} (2018) 261302}
  [\href{https://arxiv.org/abs/1901.00920}{{\ttfamily 1901.00920}}].

\bibitem{Lee:2020cfj}
S.~Lee, S.~Ahn, J.~Choi, B.~R. Ko and Y.~K. Semertzidis, \emph{{Axion Dark
  Matter Search around 6.7 $\mu$eV}},
  \href{https://doi.org/10.1103/PhysRevLett.124.101802}{\emph{Phys. Rev. Lett.}
  {\bfseries 124} (2020) 101802}
  [\href{https://arxiv.org/abs/2001.05102}{{\ttfamily 2001.05102}}].

\bibitem{Zhong:2018rsr}
{\scshape HAYSTAC} Collaboration, L.~Zhong et~al., \emph{{Results from phase 1
  of the HAYSTAC microwave cavity axion experiment}},
  \href{https://doi.org/10.1103/PhysRevD.97.092001}{\emph{Phys. Rev. D}
  {\bfseries 97} (2018) 092001}
  [\href{https://arxiv.org/abs/1803.03690}{{\ttfamily 1803.03690}}].

\bibitem{Backes:2020ajv}
{\scshape HAYSTAC} Collaboration, K.~M. Backes et~al., \emph{{A
  quantum-enhanced search for dark matter axions}},
  \href{https://doi.org/10.1038/s41586-021-03226-7}{\emph{Nature} {\bfseries
  590} (2021) 238} [\href{https://arxiv.org/abs/2008.01853}{{\ttfamily
  2008.01853}}].

\bibitem{PhysRevD.42.1297}
C.~Hagmann, P.~Sikivie, N.~S. Sullivan and D.~B. Tanner, \emph{Results from a
  search for cosmic axions},
  \href{https://doi.org/10.1103/PhysRevD.42.1297}{\emph{Phys. Rev. D}
  {\bfseries 42} (1990) 1297}.

\bibitem{McAllister:2017lkb}
B.~T. McAllister, G.~Flower, E.~N. Ivanov, M.~Goryachev, J.~Bourhill and M.~E.
  Tobar, \emph{{The ORGAN Experiment: An axion haloscope above 15 GHz}},
  \href{https://doi.org/10.1016/j.dark.2017.09.010}{\emph{Phys. Dark Univ.}
  {\bfseries 18} (2017) 67} [\href{https://arxiv.org/abs/1706.00209}{{\ttfamily
  1706.00209}}].

\bibitem{Alesini:2019ajt}
D.~Alesini et~al., \emph{{Galactic axions search with a superconducting
  resonant cavity}},
  \href{https://doi.org/10.1103/PhysRevD.99.101101}{\emph{Phys. Rev. D}
  {\bfseries 99} (2019) 101101}
  [\href{https://arxiv.org/abs/1903.06547}{{\ttfamily 1903.06547}}].

\bibitem{Schutte-Engel:2021bqm}
J.~Sch\"utte-Engel, D.~J.~E. Marsh, A.~J. Millar, A.~Sekine, F.~Chadha-Day,
  S.~Hoof, M.~Ali, K.-C. Fong, E.~Hardy and L.~\v{S}mejkal, \emph{{Axion
  Quasiparticles for Axion Dark Matter Detection}},
  \href{https://arxiv.org/abs/2102.05366}{{\ttfamily 2102.05366}}.

\bibitem{Stott:2020gjj}
M.~J. Stott, \emph{{Ultralight Bosonic Field Mass Bounds from Astrophysical
  Black Hole Spin}},  \href{https://arxiv.org/abs/2009.07206}{{\ttfamily
  2009.07206}}.

\bibitem{Arvanitaki:2010sy}
A.~Arvanitaki and S.~Dubovsky, \emph{{Exploring the String Axiverse with
  Precision Black Hole Physics}},
  \href{https://doi.org/10.1103/PhysRevD.83.044026}{\emph{Phys. Rev. D}
  {\bfseries 83} (2011) 044026}
  [\href{https://arxiv.org/abs/1004.3558}{{\ttfamily 1004.3558}}].

\bibitem{Brito:2015oca}
R.~Brito, V.~Cardoso and P.~Pani, \emph{{Superradiance}: {New Frontiers in
  Black Hole Physics}},
  \href{https://doi.org/10.1007/978-3-319-19000-6}{\emph{Lect. Notes Phys.}
  {\bfseries 906} (2015) pp.1}
  [\href{https://arxiv.org/abs/1501.06570}{{\ttfamily 1501.06570}}].

\bibitem{Ikeda:2018nhb}
T.~Ikeda, R.~Brito and V.~Cardoso, \emph{{Blasts of Light from Axions}},
  \href{https://doi.org/10.1103/PhysRevLett.122.081101}{\emph{Phys. Rev. Lett.}
  {\bfseries 122} (2019) 081101}
  [\href{https://arxiv.org/abs/1811.04950}{{\ttfamily 1811.04950}}].

\bibitem{Baryakhtar:2020gao}
M.~Baryakhtar, M.~Galanis, R.~Lasenby and O.~Simon, \emph{{Black hole
  superradiance of self-interacting scalar fields}},
  \href{https://arxiv.org/abs/2011.11646}{{\ttfamily 2011.11646}}.

\bibitem{Blas:2020nbs}
D.~Blas and S.~J. Witte, \emph{{Imprints of Axion Superradiance in the CMB}},
  \href{https://doi.org/10.1103/PhysRevD.102.103018}{\emph{Phys. Rev. D}
  {\bfseries 102} (2020) 103018}
  [\href{https://arxiv.org/abs/2009.10074}{{\ttfamily 2009.10074}}].

\bibitem{Fukuda:2019ewf}
H.~Fukuda and K.~Nakayama, \emph{{Aspects of Nonlinear Effect on Black Hole
  Superradiance}}, \href{https://doi.org/10.1007/JHEP01(2020)128}{\emph{JHEP}
  {\bfseries 01} (2020) 128}
  [\href{https://arxiv.org/abs/1910.06308}{{\ttfamily 1910.06308}}].

\bibitem{Ouellet:2018beu}
J.~L. Ouellet et~al., \emph{{First Results from ABRACADABRA-10 cm: A Search for
  Sub-$\mu$eV Axion Dark Matter}},
  \href{https://doi.org/10.1103/PhysRevLett.122.121802}{\emph{Phys. Rev. Lett.}
  {\bfseries 122} (2019) 121802}
  [\href{https://arxiv.org/abs/1810.12257}{{\ttfamily 1810.12257}}].

\bibitem{Shilon:2012te}
I.~Shilon, A.~Dudarev, H.~Silva and H.~H.~J. ten Kate, \emph{{Conceptual Design
  of a New Large Superconducting Toroid for IAXO, the New International AXion
  Observatory}}, \href{https://doi.org/10.1109/TASC.2013.2251052}{\emph{IEEE
  Trans. Appl. Supercond.} {\bfseries 23} (2013) 4500604}
  [\href{https://arxiv.org/abs/1212.4633}{{\ttfamily 1212.4633}}].

\bibitem{Marsh:2018dlj}
D.~J.~E. Marsh, K.-C. Fong, E.~W. Lentz, L.~Smejkal and M.~N. Ali,
  \emph{{Proposal to Detect Dark Matter using Axionic Topological
  Antiferromagnets}},
  \href{https://doi.org/10.1103/PhysRevLett.123.121601}{\emph{Phys. Rev. Lett.}
  {\bfseries 123} (2019) 121601}
  [\href{https://arxiv.org/abs/1807.08810}{{\ttfamily 1807.08810}}].

\bibitem{Lawson:2019brd}
M.~Lawson, A.~J. Millar, M.~Pancaldi, E.~Vitagliano and F.~Wilczek,
  \emph{{Tunable axion plasma haloscopes}},
  \href{https://doi.org/10.1103/PhysRevLett.123.141802}{\emph{Phys. Rev. Lett.}
  {\bfseries 123} (2019) 141802}
  [\href{https://arxiv.org/abs/1904.11872}{{\ttfamily 1904.11872}}].

\bibitem{TheMADMAXWorkingGroup:2016hpc}
{\scshape MADMAX Working Group} Collaboration, A.~Caldwell, G.~Dvali,
  B.~Majorovits, A.~Millar, G.~Raffelt, J.~Redondo, O.~Reimann, F.~Simon and
  F.~Steffen, \emph{{Dielectric Haloscopes: A New Way to Detect Axion Dark
  Matter}}, \href{https://doi.org/10.1103/PhysRevLett.118.091801}{\emph{Phys.
  Rev. Lett.} {\bfseries 118} (2017) 091801}
  [\href{https://arxiv.org/abs/1611.05865}{{\ttfamily 1611.05865}}].

\bibitem{Baryakhtar:2018doz}
M.~Baryakhtar, J.~Huang and R.~Lasenby, \emph{{Axion and hidden photon dark
  matter detection with multilayer optical haloscopes}},
  \href{https://doi.org/10.1103/PhysRevD.98.035006}{\emph{Phys. Rev. D}
  {\bfseries 98} (2018) 035006}
  [\href{https://arxiv.org/abs/1803.11455}{{\ttfamily 1803.11455}}].

\bibitem{Stern:2016bbw}
I.~Stern, \emph{{ADMX Status}},
  \href{https://doi.org/10.22323/1.282.0198}{\emph{PoS} {\bfseries ICHEP2016}
  (2016) 198} [\href{https://arxiv.org/abs/1612.08296}{{\ttfamily
  1612.08296}}].

\bibitem{Alesini:2017ifp}
D.~Alesini, D.~Babusci, D.~Di~Gioacchino, C.~Gatti, G.~Lamanna and C.~Ligi,
  \emph{{The KLASH Proposal}},
  \href{https://arxiv.org/abs/1707.06010}{{\ttfamily 1707.06010}}.

\bibitem{Gorghetto:2018myk}
M.~Gorghetto, E.~Hardy and G.~Villadoro, \emph{{Axions from Strings: the
  Attractive Solution}},
  \href{https://doi.org/10.1007/JHEP07(2018)151}{\emph{JHEP} {\bfseries 07}
  (2018) 151} [\href{https://arxiv.org/abs/1806.04677}{{\ttfamily
  1806.04677}}].

\bibitem{Klaer:2017qhr}
V.~B. Klaer and G.~D. Moore, \emph{{How to simulate global cosmic strings with
  large string tension}},
  \href{https://doi.org/10.1088/1475-7516/2017/10/043}{\emph{JCAP} {\bfseries
  10} (2017) 043} [\href{https://arxiv.org/abs/1707.05566}{{\ttfamily
  1707.05566}}].

\bibitem{Sikivie:2020zpn}
P.~Sikivie, \emph{{Invisible Axion Search Methods}},
  \href{https://doi.org/10.1103/RevModPhys.93.015004}{\emph{Rev. Mod. Phys.}
  {\bfseries 93} (2021) 015004}
  [\href{https://arxiv.org/abs/2003.02206}{{\ttfamily 2003.02206}}].

\bibitem{Irastorza:2018dyq}
I.~G. Irastorza and J.~Redondo, \emph{{New experimental approaches in the
  search for axion-like particles}},
  \href{https://doi.org/10.1016/j.ppnp.2018.05.003}{\emph{Prog. Part. Nucl.
  Phys.} {\bfseries 102} (2018) 89}
  [\href{https://arxiv.org/abs/1801.08127}{{\ttfamily 1801.08127}}].

\bibitem{OHare:2020wah}
C.~A.~J. O'Hare and E.~Vitagliano, \emph{{Cornering the axion with CP-violating
  interactions}},
  \href{https://doi.org/10.1103/PhysRevD.102.115026}{\emph{Phys. Rev. D}
  {\bfseries 102} (2020) 115026}
  [\href{https://arxiv.org/abs/2010.03889}{{\ttfamily 2010.03889}}].

\bibitem{Kaneta:2016wvf}
K.~Kaneta, H.-S. Lee and S.~Yun, \emph{{Portal Connecting Dark Photons and
  Axions}}, \href{https://doi.org/10.1103/PhysRevLett.118.101802}{\emph{Phys.
  Rev. Lett.} {\bfseries 118} (2017) 101802}
  [\href{https://arxiv.org/abs/1611.01466}{{\ttfamily 1611.01466}}].

\bibitem{Kalashev:2018bra}
O.~E. Kalashev, A.~Kusenko and E.~Vitagliano, \emph{{Cosmic infrared background
  excess from axionlike particles and implications for multimessenger
  observations of blazars}},
  \href{https://doi.org/10.1103/PhysRevD.99.023002}{\emph{Phys. Rev. D}
  {\bfseries 99} (2019) 023002}
  [\href{https://arxiv.org/abs/1808.05613}{{\ttfamily 1808.05613}}].

\bibitem{Arias:2020tzl}
P.~Arias, A.~Arza, J.~Jaeckel and D.~Vargas-Arancibia, \emph{{Hidden Photon
  Dark Matter Interacting via Axion-like Particles}},
  \href{https://arxiv.org/abs/2007.12585}{{\ttfamily 2007.12585}}.

\bibitem{deNiverville:2020qoo}
P.~deNiverville, H.-S. Lee and Y.-M. Lee, \emph{{New searches at the reactor
  experiments based on the dark axion portal}},
  \href{https://arxiv.org/abs/2011.03276}{{\ttfamily 2011.03276}}.

\end{thebibliography}\endgroup



\providecommand{\href}[2]{#2}\begingroup\raggedright\endgroup
 
\end{document}